\documentclass{article}
\usepackage{spconf,amsmath,graphicx,hyperref}
\usepackage{booktabs}
\usepackage{siunitx}
\usepackage{tabularx}   
\usepackage{array} 
\usepackage{makecell} 
\usepackage{xcolor}
\usepackage{multirow}
\usepackage{subcaption} 

\title{Towards Building Speech Large Language Models for Multitask Understanding in Low-Resource Languages}
\name{Mingchen Shao$^{1}$, Bingshen Mu$^{1}$, Chengyou Wang$^{1}$, Hai Li$^{2}$, Ying Yan$^{2}$, Zhonghua Fu$^{1}$, Lei Xie$^{1,*}$\thanks{* Corresponding author.}}
\address{$^{1}$Audio, Speech and Language Processing Group (ASLP@NPU), School of Computer Science,\\
         Northwestern Polytechnical University, Xi'an, China \\
         $^{2}$iQIYI, Inc., China}
\begin{document}
\ninept
\maketitle
\begin{abstract}

Speech large language models (SLLMs) built on speech encoders, adapters, and LLMs demonstrate remarkable multitask understanding performance in high-resource languages such as English and Chinese. However, their effectiveness substantially degrades in low-resource languages such as Thai. This limitation arises from three factors: (1) existing commonly used speech encoders, like the Whisper family, underperform in low-resource languages and lack support for broader spoken language understanding tasks; (2) the ASR-based alignment paradigm requires training the entire SLLM, leading to high computational cost; (3) paired speech–text data in low-resource languages is scarce. 
To overcome these challenges in the low-resource language Thai, we introduce XLSR-Thai, the first self-supervised learning (SSL) speech encoder for Thai.
It's obtained by continuously training the typical SSL XLSR model on 36,000 hours of Thai speech data. Furthermore, we propose U-Align, a speech-text alignment method that is more resource-efficient and multitask-effective than typical ASR-based alignment.
Finally, we present Thai-SUP, a pipeline for generating Thai spoken language understanding data from high-resource languages, yielding the first Thai spoken language understanding dataset over 1000 hours. 
Multiple experiments demonstrate the effectiveness of our methods in building a Thai multitask understanding SLLM. We open-source XLSR-Thai and Thai-SUP to facilitate future research.\footnote{https://huggingface.co/datasets/mcshao/Thai-understanding}
\end{abstract}
\vspace{-4.5pt}
\section{Introduction}
\vspace{-4.5pt}
\label{sec:intro}

Large language models (LLMs) have demonstrated exceptional capabilities in numerous natural language processing tasks, including text understanding, generation, and reasoning~\cite{gpt, llama, qwen2}.
This capability has promoted considerable development in speech LLMs (SLLMs), which extend the LLMs to process speech input directly.
In particular, SLLMs have shown notable success in diverse spoken language understanding tasks~\cite{qwen2audio, ding2025kimi, li2025baichuan}, including automatic speech recognition (ASR), intent classification (IC), named entity recognition (NER), and speech rephrasing (SR)~\cite{wuzhiyong,vox,lianxulisan}.

To construct SLLMs, one approach discretizes speech into tokens and trains the model with the standard next-token prediction objective~\cite{glm4voice, freeze-omni, taste}.
A more widely adopted and empirically validated paradigm leverages a pretrained speech encoder to extract continuous speech representations, which are mapped to the LLM embedding space via an adapter~\cite{salmonn, geng2024unveiling, mu2025efficient}.
Building on these designs, existing SLLMs have demonstrated remarkable performance across multiple spoken language understanding tasks in high-resource languages like English and Chinese.
However, the performance of SLLMs remains substantially constrained in low-resource languages like Thai.
To address this limitation, the research question can be summarized as: \textit{How to build SLLMs that achieve strong performance on multitask understanding in low-resource languages?}

As the core component of SLLMs for processing speech input, the speech encoder plays a vital role in capturing rich acoustic and linguistic information. 
Existing SLLMs typically use self-supervised learning (SSL) encoders or supervised ASR encoders, with the Whisper~\cite{whisper} family being a popular choice.
Although trained in large-scale multilingual speech data, their performance remains suboptimal in low-resource languages~\cite{ethai}. 
Moreover, since the Whisper family is limited to tasks such as ASR, speech translation, and voice activity detection, it imposes potential constraints on developing SLLMs for multitask understanding.

The adapter aligns the speech embeddings produced by the speech encoder with the text embedding space of the LLM, playing a crucial role in enabling the LLM to understand speech.
Existing SLLMs typically begin by optimizing only the adapter on ASR tasks within the entire SLLM framework to establish speech–text alignment, and then leverage spoken language understanding data to extend the SLLMs’ multitask understanding capabilities~\cite{wuzhiyong,salmonn,osum}.
However, since this ASR-based alignment requires training the entire SLLM to fit the ASR objective, it incurs a high computational cost, and the alignment process is restricted to the ASR target rather than establishing universal speech-text alignment.

The scarcity of multitask spoken language understanding data in low-resource languages is another critical factor limiting the performance of current SLLMs. 
Unlike ASR corpora that only require utterance-level transcriptions, multitask understanding datasets must additionally provide task-specific supervision, such as intent labels, named entity labels, and paraphrase pairs.
Since annotating speech data in such languages is costly, leveraging unlabeled data through self-supervised learning and transferring paired data from high-resource languages represent practical approaches.

\begin{figure*}[htbp]  
    \centering
    \includegraphics[clip, trim=0cm 0cm 0cm 0cm, width=0.9\linewidth]{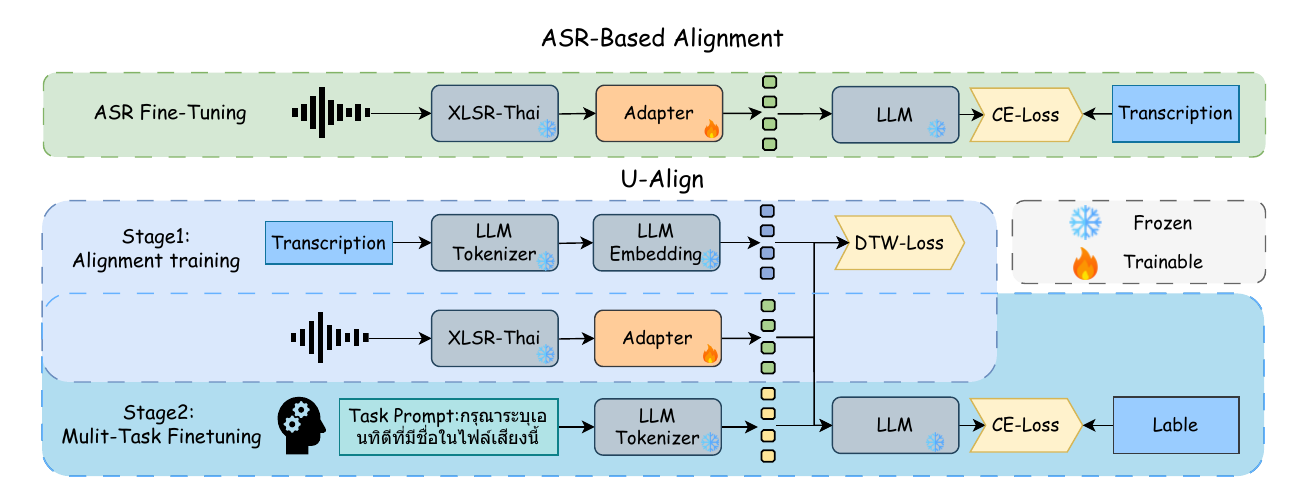}  
    \caption{\textbf{The architecture of U-Align.} Stage1: use the DTW-loss to align adapted speech representations with textual embeddings of transcriptions without involving the LLM; Stage2: initialize the adapter from Stage 1 and condition the frozen LLM with task-specific prompts and speech representations. In contrast, ASR-based alignment optimizes only the adapter on ASR tasks within the entire SLLM.}
    \vspace{-13pt}
    \label{fig:framework}
\end{figure*}
In this work, we propose a comprehensive solution for developing multitask understanding SLLMs in a low-resource language, and take Thai as a representative case.
For Thai, existing speech encoders such as the Zipformer proposed in EThai-ASR~\cite{ethai} or monsoon-Whisper-medium-gigaspeech2\footnote{\href{https://huggingface.co/scb10x/monsoon-whisper-medium-gigaspeech2}{https://huggingface.co/scb10x/monsoon-whisper-medium-gigaspeech2}} are built on limited Thai ASR annotations, and thus remain insufficient to support multitask understanding. 
Furthermore, the spoken language understanding data required for building SLLMs is entirely absent in Thai.
To leverage large amounts of unlabeled data and enhance the multitask capability of the speech encoder, we introduce XLSR-Thai, an SSL speech encoder obtained by continuously training the typical SSL XLSR model~\cite{xlsr} on 36,000 hours of Thai unlabeled speech.
Meanwhile, we propose U-Align, a more resource-efficient and multitask-effective universal speech–text alignment approach. 
Different from ASR-based alignment, which indirectly achieves speech-text alignment by optimizing the entire SLLM through the ASR task, U-Align works by directly aligning the adapted speech representations with the textual embedding of the corresponding transcriptions without involving the LLM, making the speech inputs fed into the LLM more similar to the corresponding text embeddings. 
By using this method, LLM can interpret speech as naturally as it does text, achieving a more resource-efficient and multitask-effective universal speech–text alignment. 
Besides, we propose the Thai-SUP pipeline, which generates low-resource Thai spoken language understanding data from high-resource English text understanding corpora.
This is achieved through LLM-based data augmentation and translation, followed by text-to-speech (TTS) synthesis.
Based on this pipeline, we produce the first open-source Thai spoken language understanding dataset, comprising 1,000 hours of data across IC, NER, and SR tasks.
Experimental results demonstrate that XLSR-Thai improves ASR performance and boosts multitask understanding, while U-Align achieves higher accuracy across IC, NER, SR, and ASR with lower computational cost than ASR-based alignment.

In summary, we propose a language-agnostic and transferable solution for building multitask understanding SLLMs in low-resource languages, which integrates effective encoder training, universal speech–text alignment, and data generation strategies.
Specifically, for Thai, our contributions can be outlined as follows:
\begin{itemize}
    \item \textbf{XLSR-Thai}: the first open-source Thai SSL speech encoder, providing a strong foundation for multitask understanding by extracting comprehensive speech representations.
    \item \textbf{U-Align}: a resource-efficient and multitask-effective universal speech–text alignment method that directly narrows the gap between speech representations and their corresponding text embeddings.
    \item \textbf{Thai-SUP}: a pipeline to generate low-resource spoken language understanding data from high-resource text data with LLM-based augmentation, translation, and TTS, yielding the first open-source Thai spoken language understanding dataset over 1,000 hours across IC, NER, and SR tasks.
\end{itemize}

\vspace{-4.5pt}
\section{PROPOSED METHODS}
\vspace{-4.5pt}
\label{sec:format}
To develop SLLMs with strong multitask understanding capability in low-resource languages, we propose a comprehensive solution and take Thai as a representative case. To extract rich speech representations and support multitask requirements, we continue pretraining a multilingual SSL XLSR model on readily available unlabeled speech. We further introduce U-Align, a universal speech–text alignment method that is both more resource-efficient and more effective for multitask learning. Besides, we design the Thai-SUP pipeline, which leverages LLM-based data augmentation and translation combined with TTS to transfer abundant high-resource text understanding data into low-resource spoken language understanding supervision. This approach addresses the key challenges in building low-resource language SLLMs, namely insufficient encoder capacity, suboptimal speech–text alignment, and data scarcity.

\vspace{-4.5pt}
\subsection{XLSR-Thai} 
\vspace{-4.5pt}
While speech encoders trained on ASR tasks tend to capture primarily semantic information, we first introduce the SSL speech encoder for Thai, XLSR-Thai, specifically designed to acquire both linguistic and paralinguistic cues essential for multitask understanding. Although the original XLSR model provides general speech representations from multilingual pretraining, it has seen only a few dozen hours of Thai data, leading to weak Thai-specific learning.

To address this, we develop XLSR-Thai by continuously pretraining the XLSR model on a large-scale corpus of 16,000 hours of open-source Thai speech and 20,000 hours of in-house unlabeled Thai speech. This extensive pretraining yields more robust and generalizable Thai speech representations, allowing XLSR-Thai to capture both linguistic structures and essential paralinguistic cues, making it more effective for multitask understanding.
\vspace{-4.5pt}
\subsection{U-Align}
\vspace{-4.5pt}
\subsubsection{Model architecture}
\vspace{-4.5pt}
We adopt XLSR-Thai as the speech encoder to capture both semantic and paralinguistic information. 
To bridge the speech-text modalities, we use a LayerNorm, a CNN subsampler, and a projection MLP as the modality adapter.
For the LLM decoder, we use the frozen Typhoon2-LLaMa2-3B model~\cite{typhoon}, generating text conditioned on task prompts and adapted speech embeddings.


\vspace{-4.5pt}
\subsubsection{Universal speech-text alignment}
\vspace{-4.5pt}
\label{ssec:subhead}
Traditional ASR-based alignment methods fine-tune the entire SLLM to optimize for ASR objectives, leading to high computational costs and ASR-specific optimization.
We propose U-Align, which directly aligns the adapted speech representations with the corresponding transcription representations in the LLM embedding space. 
This approach ensures that the speech inputs received by the LLM are more similar to text embeddings, facilitating a more natural interpretation of speech and enabling universal, multitask-effective speech-text alignment.
Additionally, because the alignment stage does not involve the LLM, the computational cost is significantly reduced.
To handle the length mismatch between speech and text, we align adapted speech embeddings $H=\{h_i\}_{i=1}^{I}$ to frozen LLM text embeddings $E=\{e_j\}_{j=1}^{J}$ using a cosine-distance DTW objective. 
Let
$C_{ij}=1-\frac{\langle h_i,e_j\rangle}{\|h_i\|\,\|e_j\|}$. 
The DTW-loss can be calculated as:
\begin{equation}
\mathcal{L}_{\text{DTW-loss}}
= \frac{1}{|\pi^\star|}\;
\min_{\pi\in\mathcal{P}}
\sum_{(i,j)\in \pi} C_{ij},
\end{equation}
where $\mathcal{P}$ is the set of monotonic warping paths and $\pi^\star$ is the optimal path.
Normalizing by $|\pi^\star|$ avoids sequence-length bias.

In stage2, the frozen LLM receives task-specific prompts and speech embeddings, followed by fine-tuning the SLLM on spoken language understanding data to support multitask understanding.
A key feature of U-Align is its ability to align speech embeddings directly with the corresponding transcription embeddings, enabling the LLM to interpret speech more naturally, just as it does with text.
This alignment can be achieved using various constraint functions, such as DTW-loss or CTC-loss.
Our experiments show that DTW-loss outperforms CTC-loss, and thus, we adopt DTW-loss in this work.

\begin{figure}[t] \raggedright \includegraphics[clip, width=0.95\linewidth]{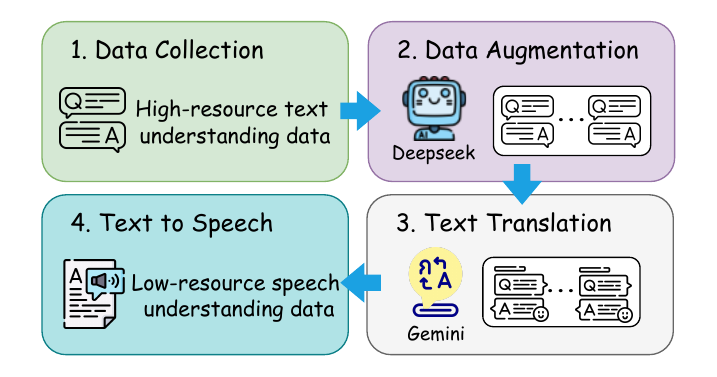} \caption{\textbf{Thai-SUP pipeline.} Thai-SUP generates low-resource Thai spoken language understanding data from high-resource English text corpora using LLM-based data augmentation, translation, and TTS.} \vspace{-9pt}\label{fig:pipe} \end{figure}
\vspace{-4.5pt}
\subsection{Thai-SUP}
\vspace{-4.5pt}
\label{ssec:subhead}
To address the scarcity of spoken language understanding data in low-resource languages, we build the Thai-SUP pipeline like Figure~\ref{fig:pipe}, which transfers supervision from high-resource text understanding corpora to low-resource spoken language understanding datasets. 
The pipeline applies LLM-based augmentation to diversify texts, translates the augmented texts into the target language, performs colloquialization and quality filtering to ensure text-to-speech (TTS) suitability, and finally synthesizes audio via TTS, thereby constructing large-scale paired speech–text supervision for spoken language understanding.

As for Thai, we start from open-source English text understanding datasets, SNIPS for IC and WikiANN / CONLL-2023 for NER. 
Each original example is augmented via DeepSeek-v3, generating ten synthetic variants per instance. 
These candidates are then filtered with Gemini-2.5-flash to remove examples that are unsuitable for downstream speech tasks. 
The remaining English examples are translated into colloquial, spoken-style Thai and rendered into speech using a Thai fine-tuned LLaSa model~\cite{llasa} to produce high-quality speech-to-text pairs.
For the SR task, we use DeepSeek-v3 to mine and select appropriate ASR speech–text pairs that lend themselves to paraphrasing, and apply Gemini-2.5-flash to generate rewritten labels.
All synthesized data yields more than 250 hours for the SR task, 648 hours for NER, and 175 hours for IC. 

\vspace{-4.5pt}
\section{EXPERIMENTS}
\vspace{-4.5pt}
\label{sec:pagestyle}

\subsection{Experimental setup}
\vspace{-4.5pt}
We continue pretraining XLSR on 16,000 hours of public Thai data, including GigaSpeech2~\cite{gigaspeech2} and MSR-86K~\cite{msr}, and 20,000 hours of in-house unlabeled Thai to obtain XLSR-Thai. 
To verify encoder gains, we fine-tune ASR on GigaSpeech2, MSR-86K, and Common Voice~\cite{commonvoice} using either XLSR-Thai or the original XLSR and report character error rate (CER). 
To assess U-Align’s effectiveness and efficiency, we compare it with a conventional ASR-based alignment under identical model settings on the same datasets. 
For multitask training, we first run U-Align’s alignment stage on a subset of 2,000 hours drawn from GigaSpeech2, MSR-86K, and Common Voice, then perform multitask fine-tuning by adding Thai-SUP to elicit multitask understanding. 
We report CER for ASR, classification accuracy (ACC) for NER and IC, and an automatic 1–5 rating for SR computed by Gemini-2.5-Flash.

\begin{table}[t]
\centering
\caption{\textbf{CER(\%) performance of XLSR-Thai.} ``Giga2 Test" indicates the Gigaspeech2 test dataset, ``CV Test" denotes the CommonVoice test dataset.}
\label{tab:xlsr}
\footnotesize  
\begin{tabular}{
    >{\raggedright\arraybackslash}p{2.7cm}  
    >{\centering\arraybackslash}p{1.3cm}  
    >{\centering\arraybackslash}p{1.3cm}  
    >{\centering\arraybackslash}p{1.3cm}    
}

\toprule
 Model & \#Params & Giga2 Test & CV Test \\
 \midrule
     Conformer-giga2 & 150M & 16.36 & 6.12  \\
                          Whisper-medium-giga2 & 769M & 14.15 & 6.92 \\
\midrule
     XLSR-AED & 450M & 17.72 & 5.73  \\
                          XLSR-Thai-AED & 450M & 14.88 & 4.80 \\
\midrule
XLSR-CTC & 300M & 16.74 & 5.06 \\
                          XLSR-Thai-CTC & 300M & \textbf{13.91} & \textbf{3.97} \\
\bottomrule
\end{tabular}
\vspace{-9pt}
\end{table}
\vspace{-4.5pt}
\subsection{Evaluation of XLSR-Thai’s effectiveness}
\vspace{-4.5pt}

\sisetup{
  detect-all,
  round-mode=places,
  round-precision=2,
  group-digits = false,
}

\newcolumntype{C}[1]{>{\centering\arraybackslash}p{#1}}

\begin{table*}[th]
\centering
\caption{\textbf{Multitask Thai spoken language understanding results.} Evaluation metrics: ACC (\%) $\uparrow$ for IC, ACC (\%) $\uparrow$ for NER (NER-ALL for overall, NER-PER for person, NER-ORG for organization, NER-LOC for location, NER-OTH for other entity types); LLM-score (1-5) $\uparrow$ for SR; CER (\%) $\downarrow$ for ASR. Directly-MT trains multitask understanding without pre-alignment.} \vspace{-9pt}
\label{tab:multi_task_results}
\footnotesize
\scalebox{0.97}{
\begin{tabular}{%
  l 
  C{1.1cm} 
  C{1.3cm} 
  C{1.4cm} 
  C{1.4cm} 
  C{1.4cm} 
  C{1.3cm} 
  C{1.1cm} 
  C{1.1cm} 
}
\toprule
\makecell[l]{ Model} 
& \makecell{IC} 
& \makecell{NER-ALL} 
& \makecell{NER-PER} 
& \makecell{NER-LOC} 
& \makecell{NER-ORG} 
& \makecell{NER-OTH} 
& \makecell{SR} 
& \makecell{ASR} \\
\midrule
Whisper + ASR-based Alignment & \num{77.15} & \num{37.86} & \num{35.61} & \num{40.83} & \num{38.29} & \num{83.27} & \num{2.66} & \num{14.43} \\
  Whisper + U-Align (DTW) & \num{81.24} & \num{42.52} & \num{43.55} & \num{47.28} & \num{40.09} & \num{87.17} & \num{2.91} & \num{14.08} \\
\midrule
XLSR-Thai + Directly-MT & \num{82.26} & \num{39.53} & \num{41.56} & \num{40.90} & \num{39.01} & \num{88.28} & \num{2.71} & \num{14.83} \\
XLSR-Thai + ASR-based Alignment & \num{81.71} & \num{43.23} & \num{47.88} & \num{46.43} & \num{41.89} & \num{87.91} & \num{2.89} & \num{13.81} \\
XLSR-Thai + U-Align (CTC) & \num{86.98} & \num{51.07} & \num{48.77} & \num{52.31} & \num{45.43} & \num{87.69} & \textbf{\num{3.10}} & \num{13.51} \\
XLSR-Thai + U-Align (DTW)      & \textbf{\num{89.68}} & \textbf{\num{53.77}} & \textbf{\num{53.92}} & \textbf{\num{54.43}} & \textbf{\num{48.09}} & \textbf{\num{90.91}} & \num{3.02} & \textbf{\num{13.32}} \\
\bottomrule
\end{tabular}}
\end{table*} 

To validate the advancement of the XLSR-Thai encoder, we conducted experiments on both ASR single-task and multitask understanding. In the ASR single-task, we fine-tuned the SSL encoder using two approaches: (i) a CTC approach, where the SSL encoder and CTC layer are fully fine-tuned, and (ii) an AED approach, where the SSL encoder is frozen and used as a feature extractor for a Conformer encoder and Transformer decoder AED model. Besides, we trained a same-size AED Conformer-giga2 model with the same data.

As shown in Table~\ref{tab:xlsr}, our XLSR-Thai outperforms the original XLSR model in both fine-tuning methods. Additionally, when compared with the Conformer-giga2 model, XLSR-Thai-AED shows significant improvements, indicating that our SSL model yields better speech representations. Furthermore, when compared with the open-source Monsoon-Whisper-Medium-GigaSpeech2, XLSR-Thai also demonstrates higher potential.

In multitask understanding, as shown in Table~\ref{tab:multi_task_results}, using XLSR-Thai consistently leads to better results than using Whisper as the encoder, both for ASR-based align and U-Align approaches. This highlights that XLSR-Thai is more effective for supporting multitask understanding in SLLM construction.
\vspace{-4.5pt}
\subsection{Validation of U-Align’s universal speech-text alignment}
\vspace{-4.5pt}
To verify that U-Align provides multitask-effective universal speech-text alignment, we conduct multitask understanding experiments. 
We design the following experiments.
\textbf{XLSR-Thai+ASR-based Alignment:} first trains modality alignment for one epoch on 2000 hours of ASR data using ASR-based alignment, then adds one epoch of multitask training with Thai-SUP, adopting XLSR-Thai as speech encoder.
\textbf{XLSR-Thai+Directly-MT:} directly trains multitask capability on ASR data combined with Thai-SUP for two epoch, without a separate alignment stage.
\textbf{XLSR-Thai+U-Align:} follows our proposed two-stage method, training one epoch of alignment with U-Align before adding Thai-SUP for multitask understanding training.
\textbf{XLSR-Thai+U-Align(CTC):} replaces the DTW-loss in the alignment stage with CTC-loss.
\textbf{Whisper+ASR-based Alignment:} replaces the encoder in XLSR-Thai+ASR-based Alignment with monsoon-Whisper-medium-GigaSpeech2.
\textbf{Whisper+U-Align:} uses the monsoon-Whisper-medium-gigaspeech2 encoder and applies U-Align for training.

The experimental results are shown in Table~\ref{tab:multi_task_results}. 
Comparing XLSR-Thai+ASR-based Alignment, XLSR-Thai+Directly-MT, and XLSR-Thai+U-Align, we observe that performing speech-text alignment before multitask understanding training yields better performance than direct multitask understanding training.
Moreover, U-Align achieves superior results over ASR-based alignment, indicating that it provides a more universal and multitask-effective alignment.
The comparison between Whisper+ASR-based alignment and Whisper+U-Align also demonstrates that U-Align consistently improves alignment across different encoders, confirming the robustness of our method.
\begin{figure}[t] \centering \includegraphics[clip, trim=0cm 0cm 0cm 0cm,width=0.85\linewidth]{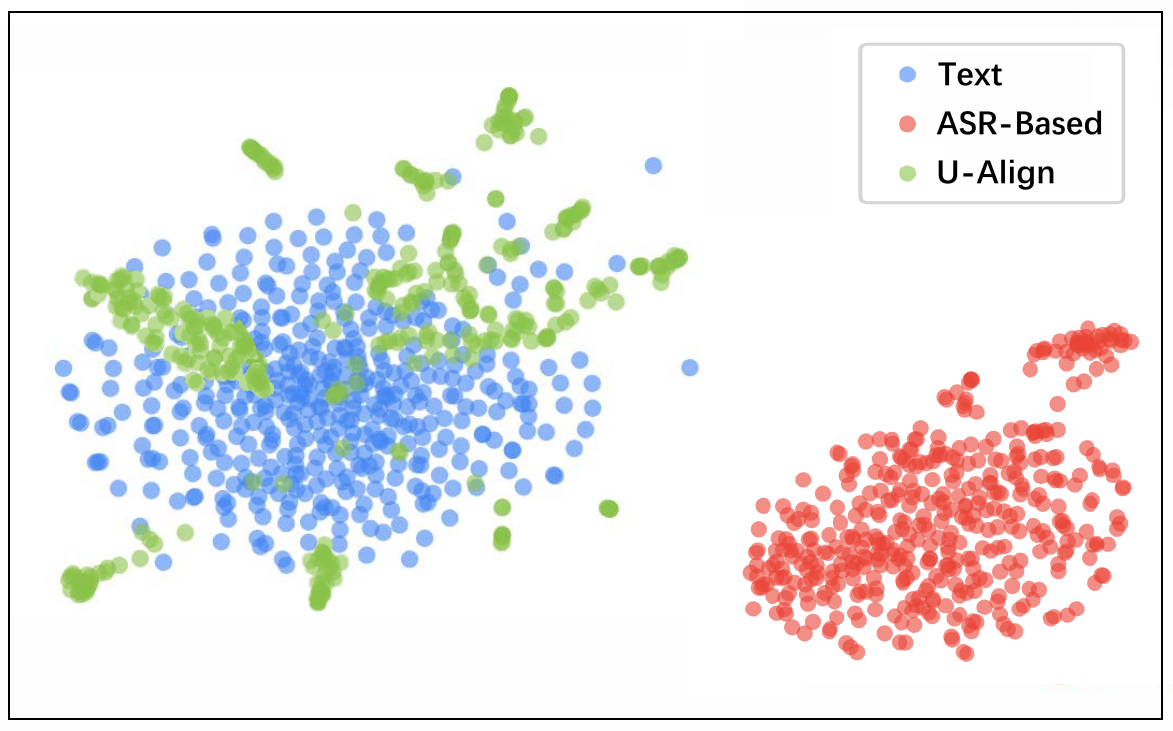} \caption{t-SNE visualization of text embedding, ASR-based embedding, and U-Align embeddings.} \vspace{-9pt} \label{fig:tsne} \end{figure}
\vspace{-4.5pt}
\subsection{Effectiveness and efficiency of U-Align}
\begin{figure}[t] \raggedright \includegraphics[clip, width=0.95\linewidth]{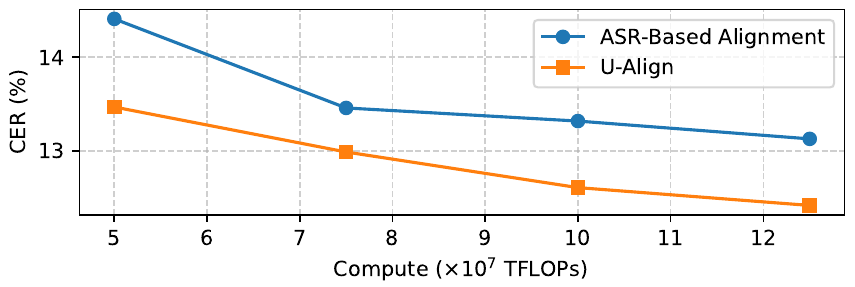} \caption{Comparison of CER(\%) performance and compute cost.} \vspace{-9pt} \label{fig:ee} \end{figure}

\vspace{-4.5pt}
We validate U-Align’s effectiveness and efficiency on the ASR task. 
The baseline trains the SLLM on ASR data with ASR-based alignment, while our method uses the same data in two stages: Stage1 learns modality alignment with U-Align, and Stage2 fine-tunes on ASR. 
We measure effectiveness by comparing the performance achieved by the models under the same computational cost, and efficiency by comparing the computational cost required to achieve the same performance. The experimental results shown in Fig. ~\ref{fig:ee}, demonstrate that U-Align consistently performs below ASR-Based Alignment, indicating that U-Align is both more efficient and more effective compared to ASR-Based Alignment.

\vspace{-4.5pt}
\subsection{Ablation study and visualization}
\vspace{-4.5pt}
As shown in Table~\ref{tab:multi_task_results}, U-Align(CTC) performs slightly worse than U-Align(DTW) but still demonstrates a significant advantage over ASR-based alignment, proving that our method is not limited to DTW-loss; any loss function that constrains speech representations and their corresponding text embeddings can be applied, and it consistently outperforms conventional ASR-based alignment.
Fig.~\ref{fig:tsne} shows t-SNE projections of speech and transcription embeddings. 
The U-Align embeddings (green) are notably fit to the Text embeddings (blue) compared to the ASR-Based embeddings (red), which are more dispersed. 
This demonstrates that U-Align aligns speech representations more closely with text, supporting its effectiveness for multitask understanding.

\vspace{-4.5pt}
\section{Conclusion}
\vspace{-4.5pt}
\label{sec:pagestyle}
In this work, we propose a comprehensive solution for building multitask understanding SLLMs for low-resource languages. 
We leverage easily accessible unlabeled data for continuously pretraining XLSR, and introduce U-Align to achieve more resource-efficient and multitask-effective speech-text alignment, and develop the Thai-SUP pipeline to transfer high-resource text understanding data to low-resource spoken language understanding data. 
Our methods are demonstrated through experiments on Thai, and this approach can be extended to any low-resource language.

\vfill\pagebreak

\bibliographystyle{IEEEbib}
\bibliography{strings,refs}

\end{document}